\documentclass[onecollarge,natbib]{svjour2}
\bibpunct{[}{]}{;}{n}{}{,} 
\smartqed  
\usepackage{graphicx}
\usepackage{epsfig}
\usepackage{graphicx,amsmath}
\newcommand\ba{\begin{eqnarray}}
\newcommand\ea{\end{eqnarray}}

\newcommand{\be}{\begin{equation}}
\newcommand{\ee}{\end{equation}}

\newcommand{\bas}{\begin{eqnarray*}}
\newcommand{\eas}{\end{eqnarray*}}

\def\sla#1{\rlap\slash #1}

\newcommand{\bno}{\begin{eqnarray*}}
\newcommand{\eno}{\end{eqnarray*}}

\def\qbar{\bar{q}}

\journalname{Few-Body Systems}
\begin{document}
\title{In-medium $\rho$-meson properties in a light-front approach
\thanks{
This work was 
supported by the Funda\c c\~ao de Amparo \`a 
Pesquisa do Estado de  S\~ao Paulo~(FAPESP),~Brazil, 
 and Conselho Nacional de Desenvolvimento Cient\'ifico e Tecnol\'ogico 
 of Brazil~(CNPq).}
 }
\author{J.~P.~B.~C.~de~Melo \and K.~Tsushima }
\institute{Laborat\'orio de F\'\i sica Te\'orica e 
Computacional - LFTC, Universidade Cruzeiro do Sul, 
01506-000 S\~ao Paulo, Brazil
\email{joao.mello@Cruzeirodosul.edu.br}             
\and
\email{ kazuo.tsushima@gmail.com, kazuo.tsushima@cruzeirodosul.edu.br}
}
\date{Received: date / Accepted: date}

\maketitle

\begin{abstract}

Properties of $\rho$-meson in symmetric nuclear matter are investigated 
within a light-front constituent quark model~(LFCQM),  
using the in-medium input calculated by the quark-meson coupling 
(QMC) model. The LFCQM used here was previously applied 
in vacuum to calculate the $\rho$-meson electromagnetic properties, namely,  
charge~$G_0$, magnetic~$G_1$, and quadrupole~$G_2$ form factors,   
as well as the electromagnetic radius and decay constant. 
We predict the in-medium modifications of the $\rho$-meson 
electromagnetic form factors in symmetric nuclear matter.


\end{abstract}

\section{Introduction}
\label{intro}

One of the main objectives in hadronic physics is to understand the  
hadron structure in terms of the fundamental degrees of freedom in  
quantum chromodynamics~(QCD), i.e, quarks and gluons. 
The Standard Model~(SM) of elementary particles includes QCD as the 
strong interaction sector. Although solving QCD is an important 
part for understanding the SM of the particles physics, 
it is still a very difficult task~(see Refs.~\cite{Ydarin,Skands2012} for QCD details).

On the other hand, light-front approach is an alternative to 
calculate observables with the ingredients from QCD~\cite{Brosky1998}. 
With light-front quantum field theory, it is possible to describe 
the lower Fock components of the hadronic bound-state wave functions, 
i.e, mesons and baryons in terms of 
quarks and gluons~\cite{Brosky1998,Jaus1990,Jaus1991,Melo2006,Lev1999,Vary2010}.

In the present work the previous model for the $\rho$-meson in vacuum~\cite{Melo1997,Melo2012b,Anace2014}  
is used to calculate the in-medium electromagnetic observables   
such as the electromagnetic form factors,~charge~$G_0$, magnetic~$G_2$, and quadrupole~$G_2$, 
as well as the electromagnetic radius and decay constant, 
using the plus-component of the current~$J^\mu$ applied in symmetric nuclear matter~\cite{Melo2014}.
(For a review of the in-medium properties of hadrons, see e.g., Refs.~\cite{Hayano2010,Brooks2011}.)
Next, we present the light-front model of $\rho$-meson, 
and briefly review the main ingredients of the model.

\section{The Light-front model}
\label{sec1}

The electromagnetic current 
in the impulse approximation for the spin-1 bound state systems, with the 
plus component of the electromagnetic current~$J^+$,~is given by, 
\begin{eqnarray}
J^+_{ji} & = & \imath \int\frac{d^4k}{(2\pi)^4}
 \frac{Tr[\epsilon^{'\alpha}_j \Gamma_{\alpha}(k,k-p_f)
(\sla{k}-\sla{p_f} +m) \gamma^{+} 
(\sla{k}-\sla{p_i}+m) \epsilon^\beta_i \Gamma_{\beta}(k,k-p_i)
(\sla{k}+m)]}
{((k-p_i)^2 - m^2+\imath\epsilon) 
(k^2 - m^2+\imath \epsilon)
((k-p_f)^2 - m^2+\imath \epsilon)}
\nonumber \\ & &\times 
\Lambda(k,p_f)\Lambda(k,p_i) \ ,
\label{electro}
\end{eqnarray}
where  $m$ is the quark mass, 
and $\epsilon^{\beta}_i$ and $\epsilon^{'\alpha}_j$ ($i,j=x,y,z$) 
are the quatri-vector polarizations, respectively, for the 
initial state $\rho$-meson,  
\begin{equation}
\epsilon^{\beta}_x=(-\sqrt{\eta},\sqrt{1+\eta},0,0),
~\epsilon^{\beta}_y=(0,0,1,0),~\epsilon^{\beta}_z=(0,0,0,1), 
\end{equation}
and for the final state $\rho$-meson,
\begin{equation}
\epsilon^{\prime \alpha }_x=(\sqrt{\eta},\sqrt{1+\eta},0,0),~\epsilon^{\prime \alpha}_y=(0,0,1,0),
~\epsilon^{\prime \alpha }_z=(0,0,0,1), 
\end{equation}
with~$\eta \equiv -q^2/4m^2_{\rho}=Q^2/4m^2_\rho > 0$~(q; four-momentum transfer).

The electromagnetic current,~Eq.~(\ref{electro}), is divergent, and in order to 
turn the Feynman amplitude finite, the regulator function~$\Lambda(k,p)$ is utilized 
~\cite{Melo1997}, 
\begin{equation}
\Lambda(k,p)=\frac{1}{((k-p)^2 - m^2_R + \imath \epsilon )^2},
\end{equation}
here, the regulator mass~$m_R$, is chosen to reproduce the experimental value of the $\rho$-meson decay 
constant in vacuum,~$156\pm8$~MeV~\cite{PDG2014}. 

The vertex with the spinor structure for $\rho$-$q\bar{q}$ is modeled by, 
\begin{equation}
\Gamma^\mu (k,k') = \gamma^\mu -\frac{m_\rho}{2}
 \frac{k^\mu+k'^\mu}{ p.k + m_\rho m -\imath \epsilon}  \ ,
\label{eq:rhov}
\end{equation}  
where, the $\rho$-meson is on-mass-shell, and its four momentum
is $p^\mu \ = \ k^\mu \ - \  k'^\mu$, the quark momenta
are $k^\mu$ and $k'^\mu$, and their masses by $m$~\cite{Melo1997,Melo2012}.

The light-front $\rho$-meson wave function is obtained after the substitution of the on-mass-shell condition, 
$k^-=(k^2_{\perp}+m^2)/k^+$ in the propagator of the quark that absorbs
the photon and in the corresponding regulator,
\begin{eqnarray}
\Phi_i(x,\vec k_\perp)=\frac{N^2}{(1-x)^2(m^2_\rho-M_0^2)
(m^2_\rho- M^2_R)^2} 
\vec \epsilon_i . [\vec \gamma -  \frac{\vec k}{\frac{M_0}{2}+ m}] \ ,
\label{eq:npwf}
\end{eqnarray}
where the polarization state is given by $\vec \epsilon_i$. The wave-function
corresponds to an S-wave state~\cite{Jaus1990}.  
The free quark-antiquark mass operator, and the $M_R$ function are 
given below, 
\begin{eqnarray}
M^2_0 & = &  \frac{k^2_\perp+m^2}{x} 
+\frac{(\vec p-\vec k)^2_\perp+m^2}{1-x}-p_\perp^2, \\ \nonumber
M^2_R & = &  \frac{k^2_\perp+m^2}{x} 
+\frac{(\vec p-\vec k)^2_\perp+m^2_R}{1-x}-p_\perp^2~,
\end{eqnarray}
with $x=k^+/p^+$.

In the case of spin-1 particles with the light-front 
approach, the matrix elements of the plus-component of the electromagnetic current,
$J^+$, is constrained by the 
angular condition~\cite{Melo1997,Cardarelli1995,Chung1988}: 
\begin{eqnarray}
 \Delta(q^2)= ( 1 + 2 \eta ) I^+_{11} + I^+_{1-1} + 
 - \sqrt{8 \eta} I^+_{10}-I^+_{00}=0~.
 \end{eqnarray}
With the equation above, it is possible to express the electromagnetic form factors 
with different linear combinations~\cite{Melo1997,Cardarelli1995}, and eliminate 
some matrix elements~$I^+_{m'm}$,~($m'=0,1;m=\pm1,0)$. 
But, some linear combinations of the electromagnetic matrix 
elements,~$J_{ji}^+$, break the covariance as well as the 
rotational symmetry, because 
the zero modes contributions or the pair terms~\cite{Melo2012,Bakker2002,Clayton2015}. 
In order to restore the covariance, it is necessary to add the pair-term contributions 
or zero modes~\cite{Melo1997, Melo2012,Bakker2002}.
The light-front and instant form spin bases are connected by the 
Melosh rotation matrix~(see Ref.~\cite{Melo1997} for 
details).

However, in the combinations of the matrix elements of the electromagnetic 
current from Grach et al.~\cite{Inna84}, the zero modes are canceled out. This was demonstrated 
numerically in Ref.~\cite{Melo1997} and 
analitically in Ref.~\cite{Melo2012},~since the 
electromagnetic matrix element of the current,~$I^+_{00}$,
was eliminated by the angular condition~\cite{Melo1997,Cardarelli1995,Inna84}. 

The electromagnetic form factors with the prescription above, are  
given below in both the light-front spin basis,~$I^+_{mm'}$, 
and the instant form spin basis,~$J^+_{ji}$:
\begin{eqnarray}
 G^{GK}_0 & = & \frac{1}{3}\left[\left( 3 - 2 \eta \right)
 I^+_{11} + 2 \sqrt{2 \eta} I^+_{11} + I^+_{1-1} \right] \nonumber \\
 & = & \frac{1}{3}\left[ J^+_{xx} + 2 J^+_{yy} - \eta J^+_{yy}  
 + \eta J^+_{zz}\right]~, \nonumber \\
 G^{GK}_1 & = & 2 \left[ I^+_{11} -\frac{1}{\sqrt{2 \eta } } I^+_{10}  \right]  \nonumber \\ 
 & = &  J^+_{yy} - J^+_{zz} -\frac{J^+_{zx}}{\sqrt{\eta}} , 
 \nonumber \\
 G^{GK}_2 & = & \frac{2 \sqrt{2}}{3} \left[ -\eta I^+_{11} + \sqrt{2 \eta} .
 I^0_{10} -I^+_{1-1}  \right] \nonumber \\
 & = & \frac{\sqrt{2}}{3} \left[ J^+_{xx} + 
 J^+_{yy} (-1 - \eta) + \eta J^+_{zz} \right]~.
\end{eqnarray}
In next section, the light-front constituent quark model described 
above is applied to study the $\rho$-meson form factors in symmetric 
nuclear matter.

\section{Rho-meson in-medium}

The light-front model for the vector $\rho$-meson was developed in Ref.~\cite{Melo1997} in order to 
calculate the observables in  vacuum. In the present work, 
we explore that model, to calculate the $\rho$-meson observables in 
symmetric nuclear matter~(see Refs.~\cite{Hayano2010,Brooks2011} 
for a review of in-medium quark and hadron properties). 
For modeling nuclear matter, "Quark-Meson Coupling"~(QMC) model 
is utilized here, and combined with the light-front approach, analogous to that 
was done for the pion case with symmetric vertex for the pion-quark coupling~\cite{Melo2014,Melo2002}.

The QMC model, which describes nuclear matter based on the quark degrees of freedom, 
was introduced using MIT bag model by Guichon~\cite{Guichon1988}, 
and Frederico et al.~\cite{Frederico1989} with a confining harmonic 
potential. The model was successfully applied for finite 
nuclei~\cite{QMCfinite}, and also mesons properties in medium~\cite{Saito2007}. 
In the QMC model, because of the surrounding medium, the 
bound state mesons and baryons are modified, compared with those in  
vacuum. For some properties calculated for symmetric nuclear matter, 
we show in Fig.~1 (negative of the binding energy),  
and summarize in table~I.
\begin{figure}[htb]
\vspace{1.0cm}
\begin{center}
\epsfig{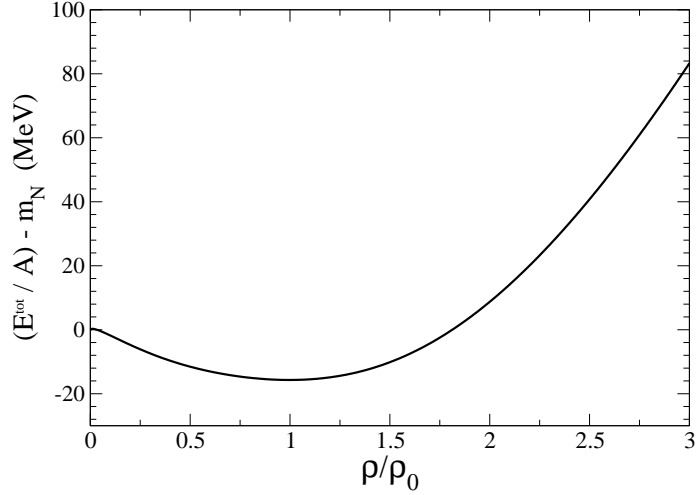}
\end{center}
\caption{\label{fig0}
Negative of the binding energy per nucleon for symmetric nuclear matter, 
calculated with the vacuum up and down quarks masses,~$m_q=430$~MeV, 
with the QMC model~\cite{Saito2007}.
}
\vspace{1.0cm}
\end{figure}

\begin{table}[ht!]
\begin{center}
\caption{The MIT bag model~(see Ref.~\cite{Saito2007} for details) 
quantities and coupling constants, the parameter $Z_N$, bag constant $B$ (in $B^{1/4}$),
and the properties for symmetric nuclear matter
at normal nuclear matter density $\rho_0 = 0.15$ fm$^{-3}$,
for $m_q = 5,~220$ and $m_q=430$~MeV. The effective nucleon mass, $m_N^*$, and the nuclear
incompressibility, $K$, are quoted in MeV (the free nucleon bag radius used is $R_N = 0.8$ fm,
the standard value in the QMC model~\cite{Saito2007}). }
\label{Tab:QMC}
\bigskip
\begin{tabular}{c|cccccc}
\hline
$m_q$(MeV)&$g_{\sigma}^2/4\pi$&$g_{\omega}^2/4\pi$
&$m_N^*$ &$K$ & $Z_N$ & $B^{1/4}$(MeV)\\
\hline
 5   &~5.39  &~5.30   &~754.6  &~279.3  &~3.295 &~170   \\
 220 &~6.40  &~7.57   &~698.6  &~320.9  &~4.327 &~148    \\
 430 &~8.73  &~11.93  &~565.3  &~361.4  &~5.497 &~69.8    \\
\hline
\hline
\end{tabular}
\end{center}
\end{table}

Next, the main ingredients of the QMC model are presented in order to calculate the 
$\rho$-meson properties in medium.  
For the system of a uniform, spin- and isospin-saturated  
symmetric nuclear matter, the effective Lagrangian density is given by~\cite{Saito2007},
\begin{equation}
{\cal L} = {\bar \psi} [i\gamma \cdot 
\partial -m_N^*({\hat \sigma}) -g_\omega {\hat \omega}^\mu \gamma_\mu ] \psi
+ {\cal L}_\textrm{meson},
\label{lag1}
\end{equation}
where ${\cal L}_{\rm meson}$ is the free meson Lagrangian, 
\begin{equation}
{\cal L}_\mathrm{meson} = \frac{1}{2} 
(\partial_\mu {\hat \sigma} \partial^\mu {\hat \sigma} - m_\sigma^2 {\hat \sigma}^2)
- \frac{1}{2} \partial_\mu {\hat \omega}_\nu (\partial^\mu {\hat \omega}^\nu - \partial^\nu {\hat \omega}^\mu)
+ \frac{1}{2} m_\omega^2 {\hat \omega}^\mu {\hat \omega}_\mu \ , \nonumber
\label{mlag1}
\end{equation}
and, $\psi$, ${\hat \sigma}$ and ${\hat \omega}$ are respectively the nucleon,
Lorentz-scalar-isoscalar $\sigma$, and Lorentz-vector-isoscalar $\omega$ 
field operators, with the effective nucleon mass defined by, 
\begin{equation}
m_N^*({\hat \sigma}) \equiv m_N - g_\sigma({\hat \sigma}) {\hat \sigma}~.
\label{efnmas}
\end{equation}
In the present work, the nuclear matter is in its rest frame. 
For symmetric nuclear matter with the Hatree mean-field approximation, 
the baryon~$(\rho)$ and scalar~($\rho_s$) densities, are calculated as,
\begin{eqnarray}
\rho &=& \frac{4}{(2\pi)^3}\int d\vec{k}\ \theta (k_F - |\vec{k}|)
= \frac{2 k_F^3}{3\pi^2},
\label{rhoB} \\ 
\rho_s &=& \frac{4}{(2\pi)^3}\int d\vec{k} \ \theta (k_F - |\vec{k}|)
\frac{m_N^*(\sigma)}{\sqrt{m_N^{* 2}(\sigma)+\vec{k}^2}},
\label{rhos}
\end{eqnarray}
In the equations above, $k_F$ is the Fermi momentum, 
and~$m^*_N(\sigma)$ is the value of the effective nucleon mass at a given  
density, self-consistently calculated with the QMC model~\cite{QMCfinite,Saito2007}.
The Dirac equations for the light quark and antiquark, are given by,
\begin{eqnarray}
\left[ i \gamma \cdot \partial_x -
(m_q - V^q_\sigma)
\mp \gamma^0
\left( V^q_\omega +
\frac{1}{2} V^q_\rho
\right) \right]
\left( \begin{array}{c} \psi_u(x)  \\
\psi_{\bar{u}}(x) \\ \end{array} \right) &=& 0,
\label{diracu} \nonumber \\
\left[ i \gamma \cdot \partial_x -
(m_q - V^q_\sigma)
\mp \gamma^0
\left( V^q_\omega -
\frac{1}{2} V^q_\rho
\right) \right]
\left( \begin{array}{c} \psi_d(x)  \\
\psi_{\bar{d}}(x) \\ \end{array} \right) &=& 0,
\label{diracd}
\end{eqnarray}
here, the Coulomb interaction is neglected, because the nuclear matter 
has the properties adequately described by the strong interactions, also, the 
SU(2) symmetry for the light quarks~($m_q=m_u=m_d)$ is assumed. Because 
of the symmetric nuclear matter in Hartree approximation,~$V^q_\rho$ is zero.

\begin{figure}[ht]
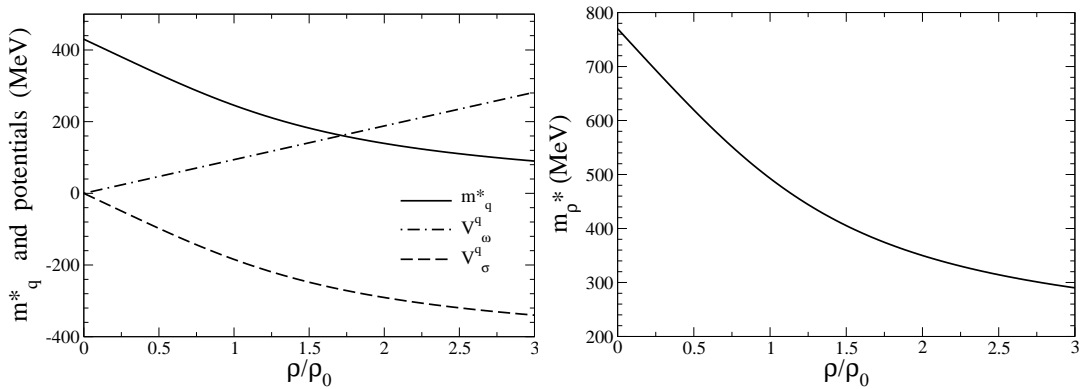

\begin{center}
\epsfig{figure=LFrho_mqstar.eps,width=7.0cm}
\epsfig{figure=LFrhorhomstar.eps ,width=7.0cm}
\end{center}
\caption{Effective constituent light-quark masses and the potentials
felt by the quarks (left panel), and 
$\rho$-meson effective mass, $m^*_{\rho}$ (right panel), in symmetric nuclear matter calculated 
by the QMC model~\cite{Saito2007}.}
\label{fig2}
\end{figure}

The bag radius in medium for the $\rho$-meson, $R_{\rho}^*$,
is determined by the 
stability condition for the mass of the hadron against the
variation of the bag radius~\cite{Saito2007} (see Eq.~(\ref{hmass})),~and, 
the eigenenergies in units of $1/R_{\rho}^*$ are given by,
\begin{eqnarray}
\left( \begin{array}{c}
\epsilon_u \\
\epsilon_{\bar{u}}
\end{array} \right)
& = &  \Omega_q^* \pm R_\rho^* \left(
V^q_\omega
+ \frac{1}{2} V^q_\rho \right),    \nonumber \\
\left( \begin{array}{c} \epsilon_d \\
\epsilon_{\bar{d}}
\end{array} \right)
& = &  \Omega_q^* \pm R_\rho^* \left(
V^q_\omega
- \frac{1}{2} V^q_\rho \right)~.
\label{energy}
\end{eqnarray}
The mass of the $\rho$-meson in medium is calculated with the expression below, 
\begin{eqnarray}
m_h^* &=& \sum_{j=q,\bar{q}}
\frac{ n_j\Omega_j^* - z_{\rho}}{R_{\rho}^*}
+ {4\over 3}\pi R_{\rho}^{* 3} B,\quad
\left. \frac{\partial m_{\rho}^*}
{\partial R_\rho}\right|_{R_\rho = R_{\rho}^*} = 0,
\label{hmass}
\end{eqnarray}
here,~$\Omega_q^*=\Omega_{\bar{q}}^*
=[x_q^2 + (R_{\rho}^* m_q^*)^2]^{1/2}$, with
$m_q^*=m_q{-}g^q_\sigma \sigma$,
and $x_{q}$ being the lowest bag eigenfrequencies; and 
$n_q$ ($n_{\qbar})$, is the light-quark (light-antiquark)  
number. 
We show in Fig.~\ref{fig2} the calculated in-medium light-quark effective mass 
and the potentials (left panel), and the effective mass of 
the $\rho$-meson (right panel) in symmetric nuclear matter.
In this study, the in-medium input calculated by the QMC model~\cite{Saito2007}, 
namely, the effective constituent quark mass $m^*_q$ and effective $\rho$-meson mass $m^*_\rho$, 
and the light-front model for the $\rho$-meson~\cite{Melo1997}, are utilized 
in order to calculate the $\rho$-meson properties in symmetric nuclear matter.

For the constituent quark model with the light-front approach 
utilized here, the sum of the constituent quark masses 
forming the bound state $\rho$-meson should be larger than the bound 
$\rho$-meson mass for the vacuum case, as well as for the case of  
the $\rho$-meson in medium, since the constituent quark and antiquark masses 
as well as the effective $\rho$-meson mass decrease 
with increasing nuclear density (see Fig.~\ref{fig2}, and expected from the QMC model, 
similar to that found in the study of the pion case~\cite{Melo2014}).
\vspace{0.32cm}

\section{Results and Conclusions}

\begin{figure}[t]
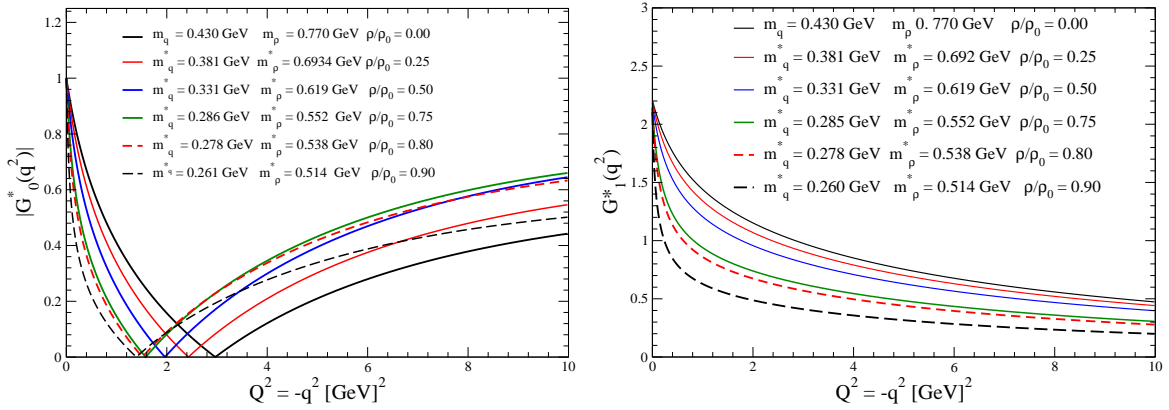

\hspace{0.85cm}
\begin{center}
\epsfig{figure=g0mediumv2.eps,width=7.5cm}
\
\epsfig{figure=g1mediumv1.eps,width=7.5cm}
\end{center}
\caption{
Electromagnetic charge form factor,~$G^*_0$~(left),
~and~~magnetic form factor,
~$G^*_1$~(right), both in symmetric nuclear matter, calculate by LFCQM~\cite{Melo1997} 
using the in-medium input obtained by the QMC model~\cite{Saito2007}.
}
\label{fig3}
\end{figure}

\begin{figure}[t]
\vspace{0.25cm}
\begin{center}
\epsfig{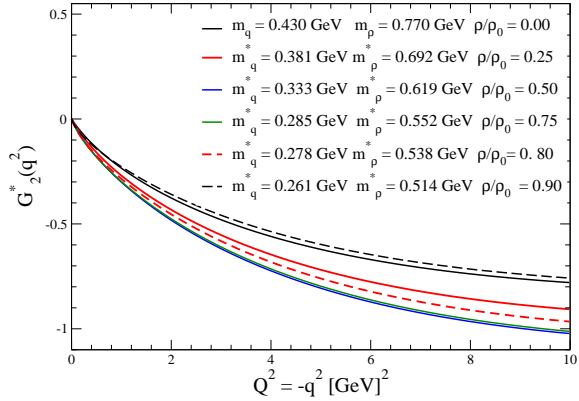}
\end{center}
\caption{
In-medium $\rho$-meson quadrupole electromagnetic form factor,~$G^*_2$,
in symmetric nuclear matter, calculate with QMC model~\cite{Saito2007} 
and LFCQM~\cite{Melo1997}.}
\label{fig4}
\end{figure}

The LFCQM model scale is obtained by adjusting the value of the 
experimental $\rho$-meson decay constant in vacuum~\cite{PDG2014,Clayton2015}, 
with the quark and antiquark mass values $m_q=m_{\bar{q}}=430$~MeV 
and the regulator mass~$m_R=3.0$~GeV. 
We show in Figs.~\ref{fig3} and~\ref{fig4} 
the calculated $\rho$-meson electromagnetic form factors, 
$G^*_0,~G_1$, and $G^*_2$ in symmetric nuclear matter.
\vspace{0.30cm}
The electromagnetic form factors,~$G^*_0,~G^*_1$ and $G^*_2$, 
are appreciably modified by the medium effects compared 
with the vacuum case in Figs.~\ref{fig3} and \ref{fig4},  
the case $\rho/\rho_0=0$ (see also ~Ref.~\cite{Melo1997}).
 
The charge electromagnetic form factor,~$G^*_0$, has a zero both in 
vacuum and in medium in the present model, which is  
about~$q^2_{zero}\simeq 3.0~GeV^3$ in vacuum~\cite{Melo1997}.
It is very interesting to notice in Fig.~\ref{fig3}~(left) that 
the zero of the charge form factor 
changes the positions for different nuclear densities due to the medium effects. 
As increasing nuclear density, the position of the zero shifts to 
the smaller $Q^2$.

The effects of nuclear medium for the magnetic moment of the $\rho$-meson  
is not so strong at $Q^2 \approx 0$. 
In vacuum, the $\rho$-meson magnetic moment $\mu$ is about~$2.20~[e/2m_\rho]$, 
and the value agrees with some recent works~\cite{Gudino2015,Simonis2016}. 
At the maximum nuclear density considered here, 
the magnetic moment has the value of about $2.10$,
which shows the decrease of $4.5~\%$ relative to the value in vacuum. 
However, the magnetic form factor decreases as increasing nuclear density 
and momentum transfer ($Q^2$), as shown in Fig.~3~(right), 
for six different nuclear densities.

Above the nuclear density~$\rho/\rho_0=0.75$, the quadrupole form factor $G^*_2$   
changes the behavior, and increases~(see Fig.~4). 
The non-zero values for the quadrupole moment 
is a consequence of the relativistic character of LFCQM. 
In addition, the reduction rate as increasing $Q^*_2$ 
is modified differently because of the medium effects.

In conclusion, we have studied the $\rho$-meson electromagnetic form factors in symmetric nuclear matter, 
charge, magnetic and quadrupole form factors up to $10~GeV^2$, with 
a relativistic constituent quark model in the light-front and using the in-medium input 
obtained by the QMC model. 
The modifications of the form factors due the medium effects are quite appreciable.
The main point of this study is to explore the effects of the medium  
with well-established model in the literature in the vacuum 
case~\cite{Melo1997}, and compare  with the results with those in symmetric nuclear matter.

\vspace{0.50cm}

{\bf Acknowledgements:} The authors would like to thank the organizers of Light Cone 2016 for invitation, and
warm hospitality, that motivated them and made them enjoy during the workshop.

 

\end{document}